\documentstyle[multicol,prb,aps,epsf]{revtex}
\begin{document}
\title{Important role  of alkali atoms in $A_4$C$_{60}$}

\author  {O. Gunnarsson}
\address{Max-Planck-Institut f\"ur Festk\"orperforschung, D-70506 Stuttgart,
Germany}

\author  {S.C. Erwin}              
\address{Complex Systems Theory Branch, Naval Research Laboratory,
Washington, D.C. 20375-5000}            

\author{E. Koch and R.M. Martin}
\address{Department of Physics, University of Illinois,
 Urbana, Illinois 61801}

 \date{\today}

\maketitle
\begin{abstract}
We    show  that     hopping via the alkali atoms plays an important
role for the $t_{1u}$ band of $A_4$C$_{60}$ (A=K, Rb), in strong contrast
to $A_3$C$_{60}$. Thus the $t_{1u}$ band is broadened by more than 40$\%$ 
by the presence of the alkali atoms.  
The difference between $A_4$C$_{60}$ and $A_3$C$_{60}$ is in particular
due to the less symmetric location of the alkali atoms in 
 $A_4$C$_{60}$.
\end{abstract}
%TCIMACRO{\TeXButton{twocolumn}{\begin{multicols{2}}}
%BeginExpansion
\begin{multicols}{2}
%EndExpansion

\section{Introduction}
The alkali doped fullerene compounds $A_3$C$_{60}$ (A=K, Rb) and
$A_4$C$_{60}$ show strikingly different properties. While $A_3$C$_{60}$ are  
metals\cite{Haddon} and superconductors,\cite{Rosseinsky,Hebard}
$A_4$C$_{60}$ are insulators.\cite{insulNMR,insulmuSR,insulPES}
The $t_{1u}$ band of  $A_4$C$_{60}$ is only partly filled, and
in contrast to $A_6$C$_{60}$, $A_4$C$_{60}$ is not a band insulator, 
since  band structure calculations predict a metal with a large
density of states at the Fermi energy.\cite{Erwininsulator,Erwinbook}
This raises interesting questions about what makes $A_4$C$_{60}$ an
insulator. To understand the  differences between 
$A_3$C$_{60}$ and $A_4$C$_{60}$, 
 it is important to understand the
electronic structure in more detail. 
It is, for instance, surprising  that for K$_4$C$_{60}$ the $t_{1u}$ 
band width (0.56 eV)\cite{Erwininsulator}
is not much smaller than for $A_3$C$_{60}$ (0.61 eV),\cite{ErwinA3C60} 
although the separation of the closest carbon atoms on neighboring
 molecules is much larger in
 $A_4$C$_{60}$ (3.5 \AA) than in $A_3$C$_{60}$ (3.1 \AA).

In $A_3$C$_{60}$ it has been demonstrated that the alkali atoms (K, Rb) 
play a small role for the states in the partly filled $t_{1u}$
 band.\cite{Saito,Martins,Satpathy}
The reason is that the alkali atoms sit in very symmetric positions
relative to the carbon atoms, and that there is a large cancellation
between different contributions to the hopping matrix elements.\cite{Satpathy}
 Here, we find that in $A_4$C$_{60}$
the indirect hopping via the alkali atoms between $t_{1u}$ orbitals 
on different C$_{60}$ molecules is of great importance, due to the
much less symmetric positions of the alkali atoms in $A_4$C$_{60}$. 
This indirect
hopping increases the $t_{1u}$ band width by more than 40$\%$.                

In Sec. II we introduce a tight-binding (TB) 
formalism\cite{Satpathy,Gunnarsson,Mele} and apply it to
$A_n$C$_{60}$ ($n=0, 3$ and 4). We show that this leads to a good
description of the $t_{1u}$ band for C$_{60}$  and $A_3$C$_{60}$
but not for $A_4$C$_{60}$. In Sec. III we include the hopping  
between the carbon and alkali atoms and show that this is important
for $A_4$C$_{60}$ but not for $A_3$C$_{60}$.
Finally, in Sec. IV we present a first-principles all-electron 
band structure calculation for the $A_4$C$_{60}$ 
structure with and without the alkali atoms, which demonstrates the 
importance of the alkali atoms.

\section{Tight-binding description of the \lowercase{$t_{1u}$} band}
To address the  differences between $A_3$C$_{60}$ and $A_4$C$_{60}$,    
we apply a TB              
formalism,\cite{Satpathy,Gunnarsson,Nozha} which was found to work well 
for $A_3$C$_{60}$.

We consider only the 60 $2p$ orbitals pointing radially out from a C$_{60}$
molecule, since these make the dominating contribution to the $t_{1u}$
band. A parametrization is then needed for the $2p-2p$ hopping 
integrals $V_{pp\sigma}$
and $V_{pp\pi}$. Following Harrison\cite{Harrison} we assume $V_{pp\pi}
/V_{pp\sigma}=-1/4$. We make the parametrization\cite{Gunnarsson}
\begin{equation}\label{eq:4}
V_{pp\sigma}=v_{\sigma}Re^{-\lambda(R-1.43)},
\end{equation}
where $R$ is the atomic separation in \AA \ and 
 $v_{\sigma}$ 
and $\lambda$ are parameters to be determined below.
 Like      Harrison\cite{Harrison}       
we only use nearest neighbor hopping inside the C$_{60}$ molecule. 
For the intermolecular hopping this prescription is ill-defined,
since many neighbors have similar separations. Since the intermolecular
separations correspond to a range where the hopping integrals decay
exponentially, we can include all hopping
integrals and still have a small contribution from distant neighbors.

The TB            Hamiltonian is solved for a free C$_{60}$ molecule 
with the bond lengths 1.40 \AA \ and 1.46 \AA. This
gives the three $t_{1u}$ orbitals
\begin{equation}\label{eq:5}
|\nu \rangle =\sum_{i=1}^{60}c_i^{\nu}|i\rangle,
\end{equation}
where $| i \rangle$ is a radial $2p$ orbital on atom $i$ and 
$c_i^{\nu}$ is   the expansion coefficient  of the $\nu$th $t_{1u}$
orbital. Using Eqs. (\ref{eq:4},\ref{eq:5}),
 we can easily calculate the hopping matrix elements between 
the $t_{1u}$ orbitals on different C$_{60}$ molecules, and obtain 
an analytical Hamiltonian describing the $t_{1u}$ band
structure.\cite{Satpathy,Nozha} 
This was applied to C$_{60}$ in the fcc structure, adjusting the
 parameters $v_{\sigma}$ to reproduce the $t_{1u}$ band width and       
$\lambda$ to reproduce the lattice parameter dependence of 
a band structure calculation\cite{Satpathy} in the local density  
approximation (LDA) of the density functional formalism.

\begin{figure}[bt]
\unitlength1cm
\begin{minipage}[t]{8.5cm}
\centerline{\epsfxsize=3.375in \epsffile{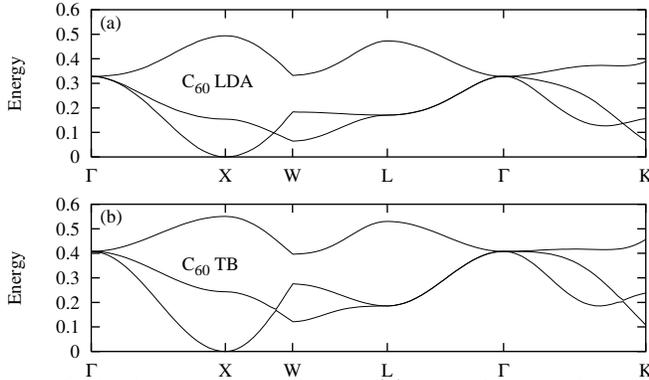}}
\caption[]{\label{fig1}Band structure for C$_{60}$ (a) according to LDA and     
(b) according to the tight-binding (TB) formalism.}
\end{minipage}
\end{figure}

In the following we compare with  LDA calculations based on a Gaussian
basis set\cite{Erwinbook} which gives slightly different band width 
than the atomic
sphere LMTO calculation in Ref. \onlinecite{Satpathy}.
The prefactor     
$v_{\sigma}=8.07$ eV/\AA \ has therefore been readjusted to reproduce the 
band width of Ref. \onlinecite{Erwinbook}. The value of $\lambda=1.98$
\AA${}^{-1}$ in Ref. \onlinecite{Satpathy} was kept unchanged. 
In Fig.~\ref{fig1} we compare the TB and LDA  
band structure calculations for C$_{60}$.\cite{Xorient}
The agreement is very good, given that only the overall band width has
been adjusted to the LDA calculation.
Within the present TB formalism, the band structure for $A_3$C$_{60}$
is the same as for C$_{60}$, apart from a small change in the band width 
due to the difference in lattice parameter. Since the band structures
of C$_{60}$ and $A_3$C$_{60}$ are very similar,\cite{Erwinbook}
it follows that our TB formalism also describes $A_3$C$_{60}$ well.
$A_3$C$_{60}$ is further discussed below.

\begin{figure}[bt]
\unitlength1cm
\begin{minipage}[t]{8.5cm}
\centerline{\epsfxsize=3.375in \epsffile{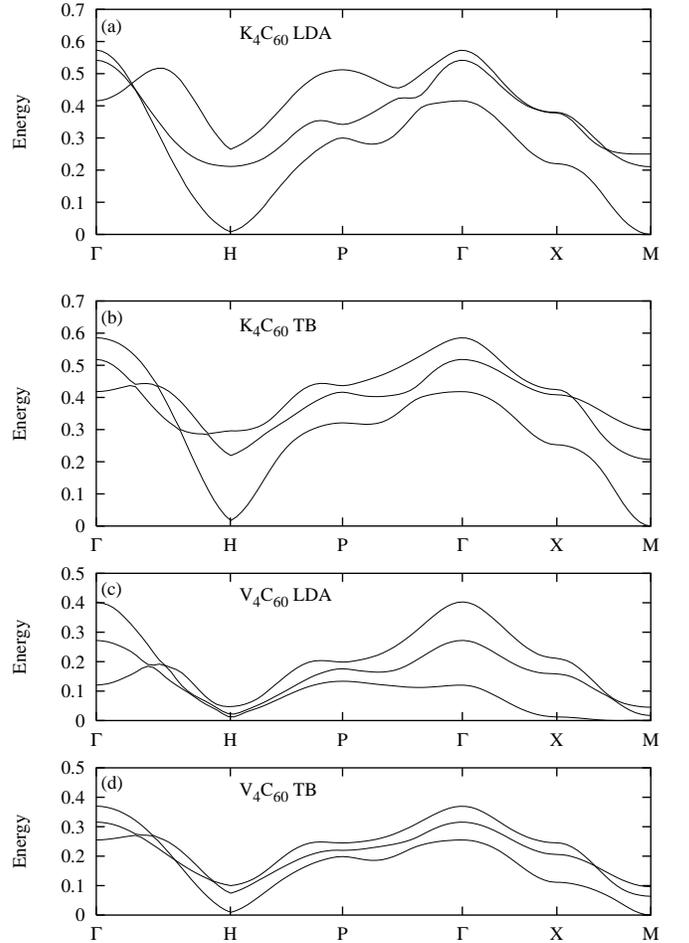}}
\caption[]{\label{fig2}The $t_{1u}$ band for K$_4$C$_{60}$ according to (a) LDA 
with alkalis,
(b) TB            with alkalis, (c) LDA           without alkalis
and (d) TB  without alkalis.}
\end{minipage}
\end{figure}

We have next    applied the same  TB formalism to K$_4$C$_{60}$.
Fig. \ref{fig2} compares the result of the TB            approach 
(d) with the LDA result (a).
There are striking differences between the results. In particular,
the TB            band-width (0.37 eV) is substantially (34$\%$)
smaller than the LDA band widths (0.56 eV). Given the difference
in the separation between the closest carbon atoms on neighboring 
C$_{60}$ molecules for C$_{60}$ (3.1 \AA) and K$_4$C$_{60}$ 
(3.5 \AA), it is, however, not surprising that the band width is 
strongly reduced compared with C$_{60}$ (0.55 eV). 
Actually, from the difference in separation alone, one would have expected
an even larger reduction (44$\%$) in the $t_{1u}$ band width 
relative to C$_{60}$. 
In the following we analyze the reason for the large deviation 
between the TB            and LDA results for K$_4$C$_{60}$
in spite of the good agreement for C$_{60}$ and $A_3$C$_{60}$.

\section{Indirect hopping via the alkali atoms}
So far the alkali atoms have been completely
neglected in the discussion. It is well-known that they indeed play
a very small role for the $t_{1u}$ band in 
$A_3$C$_{60}$.\cite{Saito,Martins,Satpathy} It was, however, observed 
by Satpathy {\it et al.}\cite{Satpathy} that the unimportance of the alkalis
is rather accidental. For instance, an alkali atom at the 
tetrahedral
position sits above a hexagon and
 has six nearest neighbor C atoms on a given C$_{60}$ molecule.
The coefficients $c_i^{\nu}$ (Eq. (\ref{eq:5})) for a given $t_{1u}$
orbital, however,  tend to change 
sign between each atom along a hexagon,\cite{Gunnarsson} 
since the $t_{1u}$ orbital is antibonding.
The result is therefore that the hopping integral between a $t_{1u}$
orbital and an alkali $s$ orbital is strongly reduced. One can therefore 
estimate\cite{Satpathy} that the mixing of $s$-character into the 
$t_{1u}$ orbital from the  tetrahedral alkali atoms is reduced
by about a factor of 22 due to cancellations between the contributions
from the different $c_i^{\nu}$'s. 
The mixing of the octahedral alkali atoms into the $t_{1u}$
orbital is even more strongly reduced.\cite{Satpathy}
It is then an interesting question
to ask if such a cancellation possibly does not take place for 
$A_4$C$_{60}$.

In $A_4$C$_{60}$ the alkali atoms are in general in less symmetric positions
relative to the neighboring C$_{60}$ molecules than in $A_3$C$_{60}$.
Relative to a given C$_{60}$
molecule, the 16 nearest neighbor alkali atoms can be grouped in 
four groups with four atoms in each. In the first group the alkali
atoms sit symmetrically above a pentagon, in the second they sit
strongly asymmetrically above a hexagon, being much closer to two of the 
hexagon atoms, in the third they sit weakly asymmetrically above a hexagon and 
in the fourth they sit strongly asymmetrically above a pentagon, being much
closer to one of the pentagon atoms. 

We now use
\begin{equation}\label{eq:6}
C=\sum_{\nu \in t_{1u}} |\sum_{i \in nn}c_i^{\nu}|^2
\end{equation}
as a measure of the coupling to a given alkali atom. The sum is over the 
$t_{1u}$ states $\nu$ and over the carbon atoms on a given 
C$_{60}$ molecule which are near neighbors           
(atoms which are less than five per cent further away than the
 nearest neighbors) of the
alkali atom considered. If we assume that the corresponding C-A hopping 
integrals are equal,
 $C$ is proportional to the square of the $t_{1u}$-A 
hopping integrals  and a measure of how much 
s-character is mixed into the $t_{1u}$ orbitals from this alkali atom.
We find that $C$ is 0.14, 0.14, 0.03 and 0.05 for an alkali
atom in the first, second third or fourth groups, respectively.
The corresponding number for each of the eight alkali atoms above a hexagon 
(tetrahedral position for $A_3$C$_{60}$) is 0.07. 
As a result, the mixing of alkali character into the $t_{1u}$ orbitals
is about a factor of $2{1 \over 2}$ larger for $A_4$C$_{60}$
than for $A_3$C$_{60}$. The large 
coupling of an alkali atom to a pentagon is due to the less drastic
cancellation between the contributions from the different atoms
than in the case of a hexagon, as has also been observed in other 
contexts.\cite{Gunnarsson} This cancellation still reduces the coupling by 
about a factor of seven for the pentagon, but much less than for
a hexagon (factor 22). In the same way, the cancellation is much less severe
in the strongly asymmetric position above a hexagon, since there are only
two nearest neighbors.   
For the weakly symmetric hexagon the cancellation is large.              
For the asymmetric pentagon, finally, the coupling to mainly just one atom,
considering only nearest neighbor interaction, is not quite enough
to give a strong coupling, although the cancellations are now less
important.

To be able to treat the C-A hopping more quantitatively, we introduce
 matrix elements which decay as the separation of the atoms 
squared.\cite{Harrison}         
To follow the spirit of Harrison, with only nearest neighbor interaction
in a system where the nearest neighbors are ill-defined,
we further introduce an exponential cut off in the matrix elements
\begin{equation}\label{eq:7}
V_{sp\sigma}=1.84D{\hbar^2\over md^2}e^{-3(d-R_{min})}
\end{equation}
where $d$ is the A-C separation and $m$ is the electron mass. 
$R_{min}$ is the shortest A-C separation for a given 
alkali atom and C$_{60}$ molecule.
The exponential cut off means that we essentially have only nearest
neighbor interaction for each A-C$_{60}$ pair.
The precise value of the factor three in the exponent is unimportant
for the following discussion, which is determined by the signs
of the coefficients $c_i^{\nu}$.
 We have introduced an adjustable parameter $D$ to be discussed below.
The indirect hopping via the alkali atoms introduces new effective 
matrix elements for the C$_{60}$-C$_{60}$ hopping
\begin{equation}\label{eq:8}
t^{\rm eff}_{\alpha \mu, \beta \nu}=\sum_{\gamma}{t_{\alpha\mu,\gamma}
t_{\beta\nu,\gamma}\over \epsilon_{t_{1u}}-\epsilon_s},
\end{equation}
where  $t_{\alpha \mu,\gamma}$ is a hopping matrix
element between the $\mu$th $t_{1u}$ orbital on the $\alpha$th C$_{60}$
molecule to the $\gamma$th alkali atom 
 and $\epsilon_{t_{1u}}$ and $\epsilon_s$ are the $t_{1u}$ and
alkali $s$ eigenvalues, respectively.
We have put $\varepsilon_s-\varepsilon_{t_{1u}}=4$ eV.
Adding these matrix elements to the direct C$_{60}$-C$_{60}$ hopping
matrix elements gives the band structure in Fig. \ref{fig2}b. The 
parameter $D$      in Eq. (\ref{eq:7}) has been set to 0.47,
which reproduces the LDA band width for K$_4$C$_{60}$.
We can see that the indirect alkali hopping increases the band width 
by about 50$\%$, although the C-A matrix elements have been reduced by
a factor of two ($1/D=2.1$) compared with Harrison's prescription.\cite{Harrison}

We next apply the same formalism to K$_3$C$_{60}$. 
K$_3$C$_{60}$ differs from C$_{60}$ due to a slightly different 
lattice parameter and the presence of the alkali atoms. The presence of 
 the alkali atoms increases
the band width by only about 6$\%$ and the largest indirect 
contribution to the 
hopping matrix elements is  down by about a factor of five compared 
with $A_4$C$_{60}$. This illustrates how much less important the 
alkali hopping is for $A_3$C$_{60}$ than $A_4$C$_{60}$.
 We consider\cite{Erwinbook} the lattice parameter 
$a=14.24$ \AA, which is somewhat larger than $a=14.2$ \AA \ used for
C$_{60}$. The increase in $a$ reduces the band width to about 0.53 eV
(from 0.55 eV for C$_{60}$), but the inclusion of the alkali 
atoms in $A_3$C$_{60}$
increases the band width again to 0.56 eV. This is still somewhat smaller than 
the width 0.61 eV found in LDA.\cite{Erwinbook}
 
The large increase in the band width for $A_4$C$_{60}$ is primarily
due to a large increase in the nearest neighbor $y-y$ hopping,
where $y$ is one of the $t_{1u}$ orbitals. The nearest neighbor indirect 
hopping goes via four alkali atoms in $A_4$C$_{60}$
but just via two atoms in $A_3$C$_{60}$. The large $y-y$ indirect hopping
in $A_4$C$_{60}$
goes via the two alkalis which are strongly asymmetric with respect to a
hexagon on one C$_{60}$ molecule and a pentagon on the other C$_{60}$
molecule. Due to the lob-structure of the $t_{1u}$ orbitals, $y-y$ hopping
via  these alkalis is very favorable, while there is no $x-x$ or $z-z$ hopping 
for symmetry reasons via these alkalis. The indirect $x-x$ hopping
 instead takes place
via alkalis located over a pentagon on one molecule and weakly 
asymmetrically over a hexagon on another molecule. The lobe structure 
is, however, less favorable for this hopping. The indirect 
nearest neighbor $z-z$ hopping is for 
symmetry reasons suppressed for hopping over all four alkalis which are
the common nearest neighbors of two neighboring C$_{60}$ molecules. 
Finally, there is efficient indirect $z-z$ hopping to the second nearest
neighbor in the $z$-direction and $y-y$ hopping to the second nearest 
neighbor in the $x$-direction. 
We observe that it is not surprising to find a large indirect second nearest
neighbor hopping via the alkalis, since the larger molecular separation
does not inhibit this hopping. 
In addition to this large indirect hopping, there is also a large
direct second nearest neighbor hopping for $A_4$C$_{60}$. The second nearest
neighbor molecules in the $z$-direction are unusually close together 
due to the compression of the bct lattice in the $z$-direction.
In addition, the coefficients $c_i^{\nu}$ in Eq. (\ref{eq:5}) 
are unuasally favorable for the second nearest neighbor hopping. 
In the present TB formalism, 
the largest total (direct plus indirect) second nearest neighbor 
hopping is therefore more than 70$\%$ of the largest 
nearest neighbor hopping, 
i.e., unusually large.

The diagonal indirect $t^{\rm eff}_{\alpha \mu,\beta\mu}$ and direct
hopping terms have the same sign as for 
both $A_3$C$_{60}$ and $A_4$C$_{60}$, while
the nondiagonal terms  have the same sign for $A_4$C$_{60}$
but different signs for $A_3$C$_{60}$.   
In the present parameter range, the band width is determined by the
diagonal elements. The differences in signs for the nondiagonal terms 
therefore do not 
influence     the band width, but they do increase 
the second moment for $A_4$C$_{60}$.    
Thus the increased dispersion       can be traced to three effects;
(i) a general increase of the alkali hopping due to the less symmetric
positions of the alkali atoms relative to the C$_{60}$ molecules,
(ii) the particularly large increase of a certain    matrix element
crucial for the band width and (iii) constructive interference of the
direct and indirect contributions to the off-diagonal matrix elements,
leading to an additional increase of the second moment.

\section{Comparison between $A_4$C$_{60}$ and $V_4$C$_{60}$}

To test these considerations, we have performed a LDA calculation for 
``$V_4$C$_{60}$,'' where the C$_{60}$ molecules are 
in exactly the same positions as for K$_4$C$_{60}$ but where the             
the potassium atoms are missing (``vacancies'').         
The calculational method was the same as described in Ref. \onlinecite{ErwinLCAO}.
The results are shown in 
Fig. \ref{fig2}c. Comparison with the K$_4$C$_{60}$ calculation
in Fig. \ref{fig2}a 
illustrates that the $t_{1u}$ band is indeed substantially narrower 
in the absence of the alkali atoms, and that the inclusion of these       
atoms increases the band width by more than 40$\%$.  This 
emphasises the importance of the alkali atoms for $A_4$C$_{60}$. 

The TB  $V_4$C$_{60}$ width (0.37 eV) is close to 
the LDA           result      (0.39 eV).
The shapes of the TB  and LDA $V_4$C$_{60}$ bands also agree rather well,     
the main difference
being a too small TB splitting at the $\Gamma$-point.
This suggests that the TB            formalism describes the C-C hopping
rather well in all three structures studied. 

\section{Summary}
To summarize, we have illustrated that the alkali atoms have a large influence
on the $t_{1u}$ band in $A_4$C$_{60}$, contrary to their small influence
on $A_3$C$_{60}$. 
The reason is, in particular, the nonsymmetric positions of the alkali atoms
 relative to the carbon atoms in $A_4$C$_{60}$, which leads to a less
efficient cancellation of the contributions to the 
$t_{1u}$-alkali hopping integrals than in
$A_3$C$_{60}$.
\section*{Acknowledgements}
We would like to thank O.K. Andersen and O. Jepsen for many helpful
discussions. This work was funded in part by the Office of Naval
Research (S.C.E.).
%and K. R\"ossmann for help with the figures.

\end{multicols}

\begin{thebibliography}{*}
\bibitem{Xorient}The C$_{60}$ molecule has been oriented so that 
the double bond with the largest $z$ coordinate is along the $x$-axis.
The band structure therefore differs from Ref. \onlinecite{Satpathy} where
the orientation is along the $y$-axis.
\end{thebibliography}
\end{document}